\documentclass[lettersize,journal]{IEEEtran}
\usepackage{amsmath,amsfonts}
\usepackage{algorithmic}
\usepackage{algorithm}
\usepackage{array}
\usepackage[caption=false,font=normalsize,labelfont=sf,textfont=sf]{subfig}
\usepackage{textcomp}
\usepackage{stfloats}
\usepackage{url}
\usepackage{verbatim}
\usepackage{graphicx}
\usepackage{cite}
\usepackage{multirow,colortbl}
\definecolor{color1}{rgb}{0.984,0.894,0.835}
\definecolor{color2}{rgb}{0.886,0.937,0.851}
\begin{document}

\begin{titlepage}
{\bfseries\LARGE IEEE Copyright Notice\par}
\vspace{1cm}
{\scshape\Large © 2022 IEEE.  Personal use of this material is permitted.  Permission from IEEE must be obtained for all other uses, in any current or future media, including reprinting/republishing this material for advertising or promotional purposes, creating new collective works, for resale or redistribution to servers or lists, or reuse of any copyrighted component of this work in other works. \par}
\vspace{1cm}
{\Large This is the author's version of an article that has been published in this journal. Changes were made to this version by the publisher prior to publication. \par}
\vspace{1cm}
{\Large The final version of record is available at http://dx.doi.org/10.1109/TIM.2022.3196949 \par}
\end{titlepage}

\title{Antenna system for trilateral drone precise vertical landing}

\author{V\'ictor Ara\~na-Pulido,~\IEEEmembership{Member,~IEEE,}
        Eugenio Jim\'enez-Ygu\'acel,~\IEEEmembership{Member,~IEEE,}
        Francisco Cabrera-Almeida,~\IEEEmembership{Member,~IEEE,}
        Pedro Quintana-Morales
\thanks{The authors are with the Institute for Technological Development and Innovation in Communications (IDeTIC), 
Department of Signals and Communication, University of Las Palmas de Gran Canaria (ULPGC), 35017 Las Palmas, Spain
(e-mail: victor.arana@ulpgc.es; eugenio.jimenez@ulpgc.es; francisco.cabrera@ulpgc.es; pedro.quintana@ulpgc.es).}
\thanks{This work was supported by the Spanish Government under Grant TEC2017-88242-C3-3-R and PID2020-116569RB-C32 Projects.}
\thanks{Manuscript received April 19, 2021; revised August 16, 2021.}}

\markboth{Journal of \LaTeX\ Class Files,~Vol.~14, No.~8, August~2021}%
{Shell \MakeLowercase{\textit{et al.}}: A Sample Article Using IEEEtran.cls for IEEE Journals}

\IEEEpubid{0000--0000/00\$00.00~\copyright~2021 IEEE}

\maketitle

\begin{abstract}
This paper presents a radio frequency system that can be used to perform precise vertical landings of drones. The system is based on the three-way phase shift detection of a signal transmitted from the landing point. The antenna system is designed by taking into account parameters such as landing tracking area, analog to digital converter (ADC) resolution, phase detector output range, antenna polarization and the effect of antenna axial ratio. The fabricated prototype consists of a landing point antenna that transmits a signal at 2.46 GHz, as well as a drone tri-antenna system that includes a phase shift detection circuitry, ADC and a simple control program that provides the correction instructions for landing. The prototype provides an averaged output data rate (ODR) suitable for landing maneuvers ($>$300 Hz). A simple system calibration procedure (detector output zeroing) is performed by aligning the antenna system. The measurements performed at different altitudes demonstrate both the correct operation of the proposed solution and its viability as an instrument for precision vertical landings. 
\end{abstract}

\begin{IEEEkeywords}
Drones, multirotors, vertical landing instruments, phase detector, tri-antenna array, ADC, averaged ODR.
\end{IEEEkeywords}

\section{Introduction}
\IEEEPARstart{P}{recision} landing systems for multirotors are especially useful when the drone must operate autonomously to recharge its batteries, approach to recharge sensors by induction, take samples of a specific terrain location or simply maneuver in an area with numerous obstacles (wooded areas, rough terrain, etc.) \cite{Woo2017, Basha2015}.

In flight stabilization maneuvers, the sensors in charge of providing pitch and roll information must provide data with a high Output Data Rate (ODR) \cite{Hoflinger2012, Garcia2017}. Thus, times of 10 or 20 ms (100 or 50 Hz) are recommended for the averaged information provided by the inertial flight system that enables the horizontal stabilization of the multirotor flight. 
The landing maneuver is one of the most complex in flight. In addition to the turbulence generated by the propellers as they beat the air over the ground, eddies of air can also be produced by the wind.
Landing options tha can compensate for the effect of these air turbulences include having pointing instructions with a sufficiently high ODR \cite{Gautam2014}.

Currently, image processing based systems are used, which are slow and require powerful processing platforms \cite{Patruno2019, Pluckter2018}. However, they have the advantage of not requiring more than a camera pointing to the $LP$, so they can be especially suitable for correction maneuvers when the drone is at a sufficient altitude and out of collision risk due to air turbulence.
GPS and the RTK variant \cite{Supej2014,Koo2017}, are frequently used in clear areas, where the positioning accuracy is high. However, the ODR is low (in the order of 1 Hz) and drops further when there are obstacles surrounding the landing point \cite{Meng2009, Dyukov2016}.
Laser tracking based system have the disadvantage of depending on beam width \cite{Lee2019}. The narrower the beam, the greater the landing accuracy, but the probability of aiming loss increases. On the contrary, if the beam width is increased to improve the pointing and to perform the landing tracking, accuracy will be lost.

Recently, a novel solution based on trilateration performed from the drone has been proposed, where only a triple receiver of the radio frequency (RF) signal transmitted from the Landing Point ($LP$) is involved \cite{Arana2021}.
The proposed system would be able to track on an inverted conical volume, where the vertex would be the $LP$ and tracking area would increase with height.
The solution was designed to provide voltages proportional to the pointing misalignment and, therefore, would make it possible to obtain a very high ODR, comparable to those currently obtained for horizontal flight stabilization. The evaluation was performed by introducing RF signals from a set of phase-synchronized generators. However, the design does not contain the necessary elements to function in real environments. It is unknown whether the inclusion of such elements must meet any specific specification or whether they may pose a significant limitation in the proper functioning of the intended application,  that is, drone precise vertical landing.

This paper addresses the problems derived from a practical realization of the landing system proposed in \cite{Arana2021}, adding and modifying those hardware and software aspects that enable an evaluation of the prototype as a precision vertical landing instrument. 
First, the most relevant aspects of the precision landing circuit presented in \cite{Arana2021} are reviewed. Then, the antenna system design is carried out, including both the onboard drone receiving tri-antenna and the landing point transmitting antenna, taking into account parameters such as: the distance between antennas, tracking area, ADC bits number, antenna polarization, axial ratio and phase shift detector, among others. Section IV describes the control system and the integration between landing system components, ending with the experimental evaluation of the designed prototype in section V.

\section{Triangular Phase Shift Landing System}
\begin{figure}[!t]
\centering
\subfloat[]{\includegraphics[width=6.5cm]{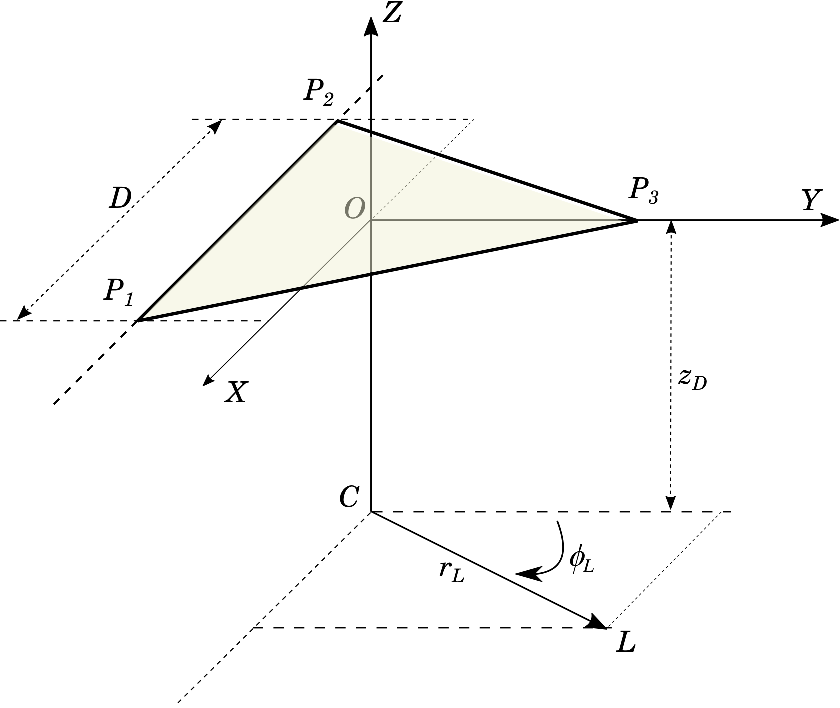}
\label{fig:figura1a}}
\hspace{0.0cm}
\subfloat[]{\includegraphics[width=8cm]{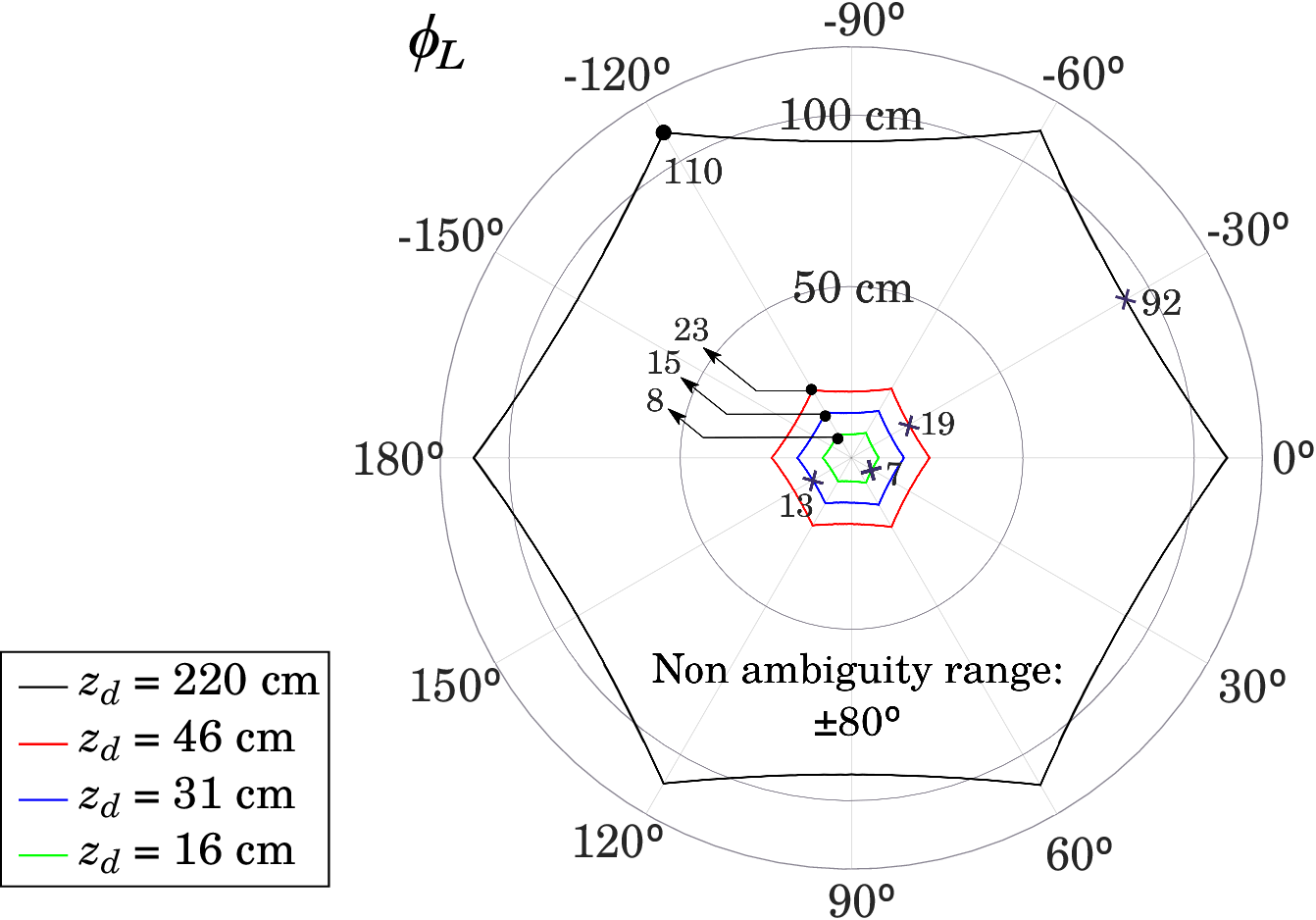}
\label{fig:figura1b}}
\caption{\textbf{(a)} Coordinate system of the drone and $LP$ reception points. \textbf{(b)} Tracking area when the drone altitude is 220, 46, 31 and 16 cm, the receiving points are separated $D$ = 7 cm, $f$ = 2.46 GHz and non-ambiguous range of $\pm$80\textordmasculine\ in the multiplier analog detector is considered.}
\label{fig:figura1}
\end{figure}

Assuming the drone is stabilized, parallel to the ground, the solution proposed in \cite{Arana2021} provides flight corrections based on the phase shift measurement between three reception points located on the drone, when a signal transmitted from the $LP$ is received (Fig. \ref{fig:figura1a}).

The triangular system is based on the assumption that the receiving points are located on an equilateral triangle with sides $D$. The non ambiguous range of the phase detector determines the maximum distance at which flight corrections can be performed correctly. This theoretical range is $\pm$90\textordmasculine\ in a multiplier analog detector but a practical range of $\pm$80\textordmasculine\ is considered because the curve detector is shifted due to mismatching.
Equations (\ref{eq:OP}) determine the reception coordinates and equations (\ref{eq:deltad}) determine differences in distance that exist between the different reception points (antennas) when a RF signal is transmitted from $LP$. The maximum theoretical phase shift in equation (\ref{eq:deltat}) ($\pm \pi$/2) gives the tracking area (${r_L}_{max}$ versus $\phi_L$) versus drone altitude ($z_D$) and considers a distance $D$ between receptions points.

The maximum distances occur when $LP$ is located at multiples of 60\textordmasculine\ ($\phi_L$ = n\,$\cdot$\,60\textordmasculine) and the minimum ones at 30\textordmasculine\,+\,n\,$\cdot$\,60\textordmasculine\ (Fig. \ref{fig:figura1b}).

\begin{eqnarray}
\label{eq:OP}
\overline{OP_1} & = &\frac{D}{2}\hat{x} - \frac{D}{2}\tan\frac{\pi}{6}\hat{y} \nonumber\\
\overline{OP_2} & = & \frac{-D}{2}\hat{x} - \frac{D}{2}\tan\frac{\pi}{6}\hat{y}\\
\overline{OP_3} & = & 0\hat{x} + \frac{D}{2\cos(\pi/6)}\hat{y} \nonumber
\end{eqnarray}

\begin{eqnarray}
\label{eq:deltad}
 \Delta d_{ij} & = & | \overline{OL} - \overline{OP_i} | - | \overline{OL} - \overline{OP_j} |\thickspace_{(ij=12,23,31)} \\
\label{eq:deltat}
 \Delta \theta_{ij} & = & 2 \cdot f \cdot \Delta d_{ij} \cdot 180 / c \quad [deg]
\end{eqnarray} 

\begin{figure}[!t]
\centering
\includegraphics[width=7.7cm]{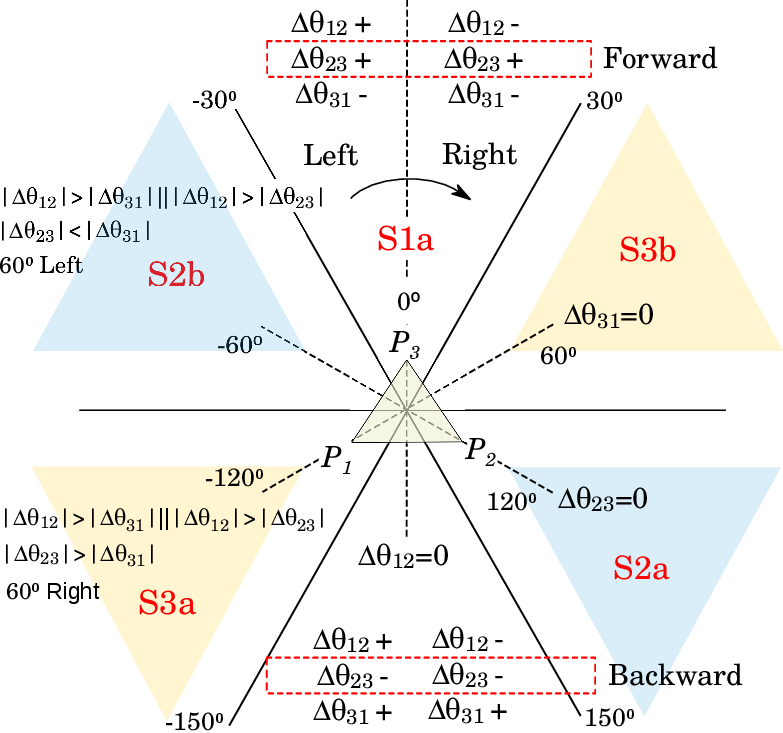}
\caption{Correction maneuvers based on the phase shift information between drone receiving points.}
\label{fig:figura2}
\end{figure}

The phase shifts between inputs ($\Delta \theta_{ij}$) and their relationship with the $LP$ location relative to drone ($r_L$ = cte \& $\phi_L$ = 0:360\textordmasculine), yield a tracking area division into 6 sectors ($S\,i\,m$ with $i$ = $1$, $2$, $3$ and $m$ = $a$, $b$ in Fig. \ref{fig:figura2}). The objective is to center the drone over $LP$, i.e., that the phase shifts between inputs be zero (same distance between each input and $LP$). For this purpose, flight maneuvers are carried out according to the signs and amplitudes of the phase shifts. First, the maneuver that leads from sector $2$ or sector $3$ to sector $1$ (60\textordmasculine\ Left or 60\textordmasculine\ Right) is performed and then the drone is guided towards $LP$. It is assumed that drone movements are right/left rotation about the $Z$-axis and forward/backward about the $OP_3$ direction (Fig. \ref{fig:figura1a}).

\section{Antenna System Design}
The antenna system consists of two antennas: one antenna located on the $LP$ that transmits the signal and another tri-antenna array ($P_1$, $P_2$ and $P_3$) located on the drone that receives the signal. The distance between antennas of the tri-antenna array ($D$) determines the area where the landing tracking can be performed (Fig. \ref{fig:figura1b}).

The voltages from phase detectors will be digitized through an ADC. This ADC must have enough resolution to be able to detect small phase shift variations that will be used to correct the flight in a continuous way.

In addition, the antenna system must ensure correct operation regardless of the turning and forward maneuvers performed by the drone. It provides unambiguous information to correct the flight and provides an optimal distance range for a given power transmitted through the LP antenna. The following sections detail the design procedure.

\subsection{Tracking area, ADC resolution and distance between antennas}
\begin{figure}[!t]
\centering
\includegraphics[width=8.3cm]{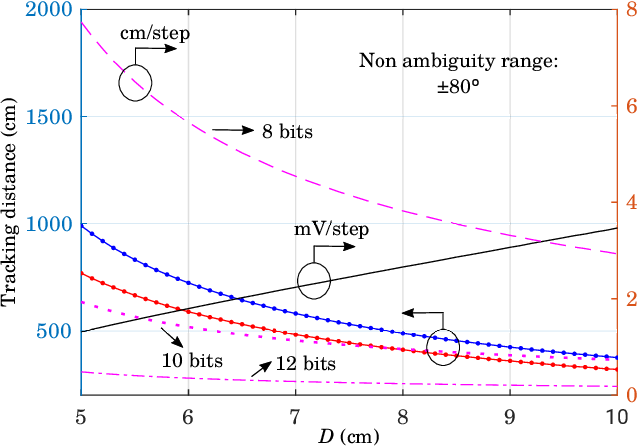}
\caption{Array design parameters as a function of the distance between receiving points ($D$): Maximum (blue) and minimum (red) tracking distance, detector sensitivity in mV/cm (black) and ADC sensitivity in cm/step (cyan) for 8, 10 and 12 bit resolution ($\Delta V_d$ = 2.6 V, $f$= 2.46 GHz, $z_D$ = 10 m and $D$ = 7 cm).}
\label{fig:figura3}
\end{figure}

From equation (\ref{eq:deltat}) and imposing the maximum non-ambiguous phase shift condition of $\pm$80\textordmasculine\ at a frequency of 2.46 GHz, the red and blue curves in Fig. \ref{fig:figura3} are obtained. These represent the minimum ($\phi_L$ = 30\textordmasculine\,+\,n\,$\cdot$\,60\textordmasculine) and maximum ($\phi_L$ = n\,$\cdot$\,60\textordmasculine) distance as a function of the separation of the receiving points ($D$) when the drone is at 10 m ($z_D$ = 10 m). Moreover, the sensitivity in terms of mV/cm (black curve) has been included. It is obtained by considering the maximum distance (2\,$\cdot\,{r_L}_{max}$) and the maximum variation of the phase detector ($\Delta V_d$) for $\pm$80\textordmasculine.

For example, if an 8-bit ADC (256 conversion steps) is available and the maximum input voltage is 5 V, a resolution of 20 mV/step is obtained. If the phase detector provides a maximum variation of $\pm$1.3 V ($\Delta V_d$ = 2.6 V) at $\pm$80\textordmasculine, the minimum distance the drone must move to produce a level change is obtained, i.e. travel distance to execute a new trajectory correction. In Fig. \ref{fig:figura3}, the minimum distance that produces a change in conversion level for 8, 10 and 12 bits when the drone altitude is 10 m has been plotted in magenta.

Finally, taking into account the available devices, a tracking distance of 7cm between receiving points ($D$ = 7 cm) has been selected. This would provide a minimum/maximum tracking distance of 419/499 cm at 10 m altitude ($z_D$ = 10 m), a 10-bit ADC over 5 V and a detector output set to $\pm$1.3 V (2.605 mV/cm and 0.746 cm/step).

\subsection{Antenna polarizations}
The analog phase detector output voltage depends on the amplitudes of the signals at its input \cite{Marcellis2015}. Therefore, the Antenna system for trilateral precise vertical landing must ensure that the amplitude received in the phase detection system does not strongly depend on the distance between transmitter and receiver nor on the drone rotation angle.

The distance dependence amplitude problem can be mitigated in the phase detectors by amplitude compression circuits such as limiters or logarithmic amplifiers, like those included in commercial circuits (AD8302) used in \cite{Arana2021, Mlynek2017, Lopez2020}. Even so, the amplitude variation resulting from the antenna system would reduce the input dynamic range and, therefore, the maximum range for tracking. This is because the range would be fixed by the drone rotation angle that provides minimum amplitude, i.e., the worst case.

\begin{figure}[!t]
\centering
\subfloat[]{\includegraphics[width=4.0cm]{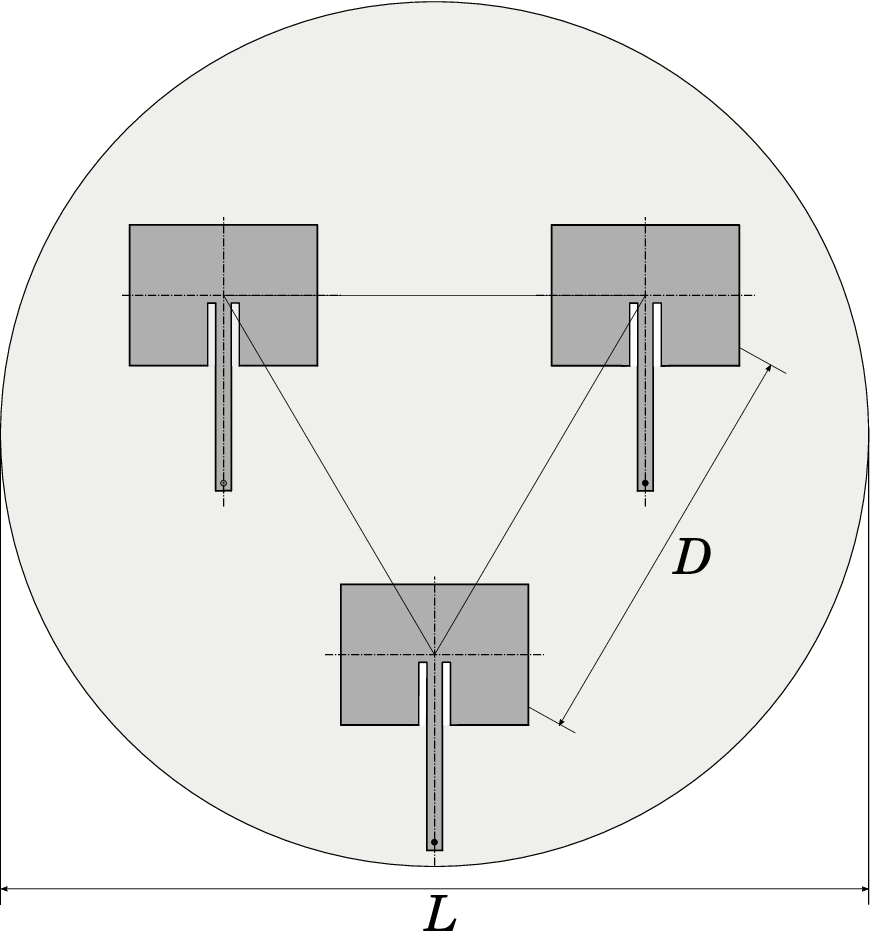}%
\label{fig:figura4a}}
\hspace{0.0cm}
\subfloat[]{\includegraphics[width=4.6cm]{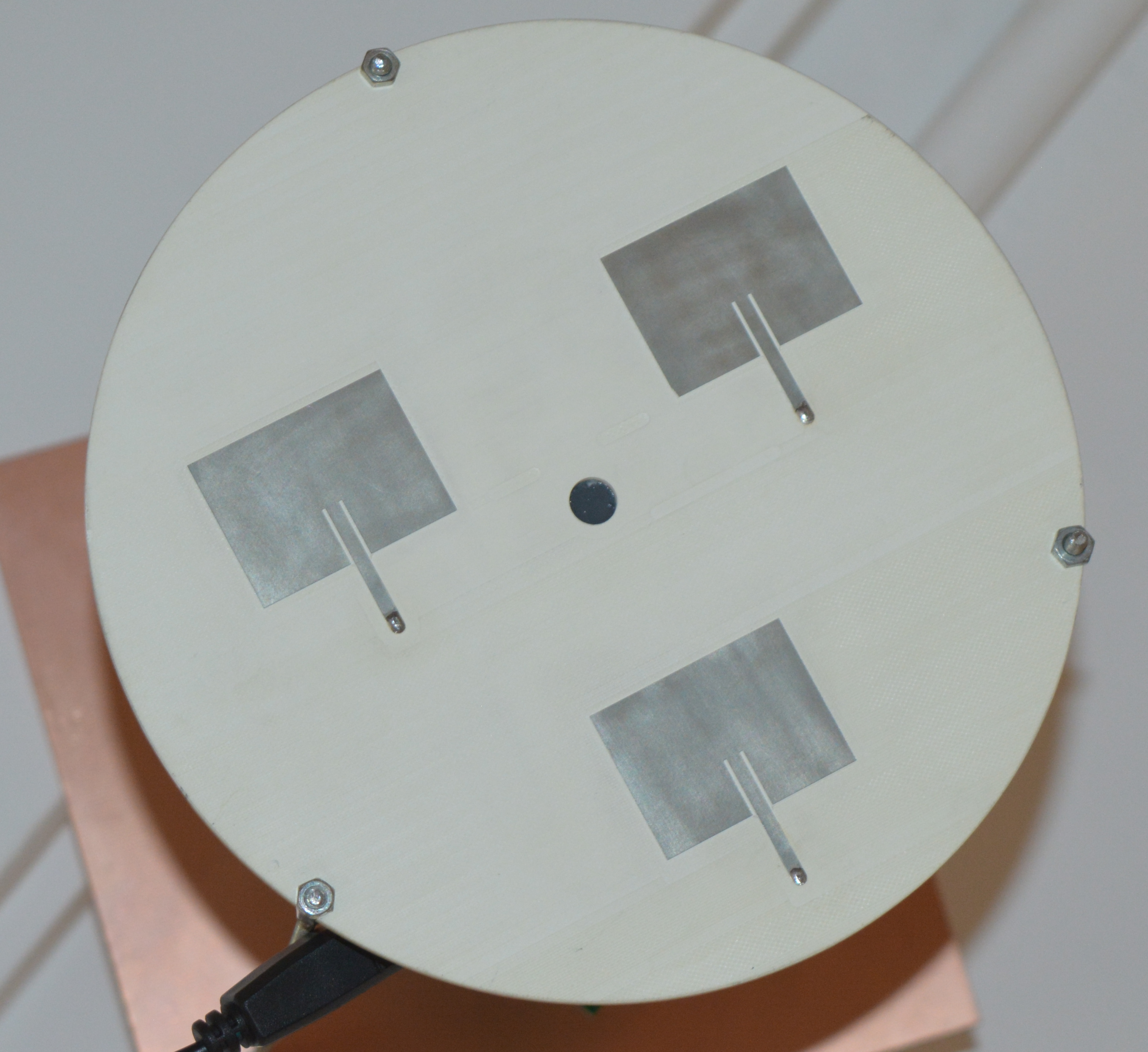}%
\label{fig:figura4b}}
\caption{Tri-antena array for drone landing system: $L$ = 15 cm and $D$ = 7 cm.}
\label{fig:figura4}
\end{figure}

To avoid amplitude variations in the received signal caused by drone rotation, a circularly polarized vs. linearly polarized schema will be used. In the $LP$, one circularly polarized patch antenna is used. Concerning the drone tri-antenna, three linearly polarized patch antennas with the same orientation are used (Fig. \ref{fig:figura4}). With that set up, the phase differences between output voltages will be proportional only to the path difference between the transmitting antenna and each of the three receiving antennas.

\subsection{Anntena system}
\begin{figure}[!t]
\centering
\subfloat[]{\includegraphics[width=8.5cm]{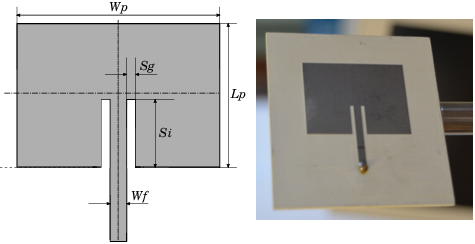}%
\label{fig:figura5a}}
\hspace{0.0cm}
\subfloat[]{\includegraphics[width=8.5cm]{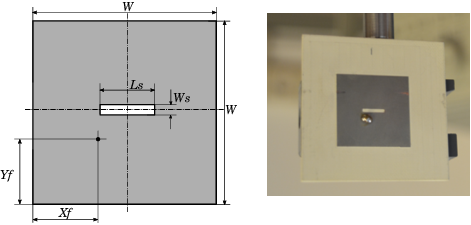}%
\label{fig:figura5b}}
\caption{\textbf{(a)} Drone and $LP$ antennas. Dimensions in mm: Linearly polarized antenna: $W\!p$ = 32.4, $Lp$ = 24.2, $Si$ = 9.2, $Sg$ = 0.6, $W\!f$ = 2.2. \textbf{(b)} Circularly polarized antenna: $W$ = 23.5, $Ws$ = 1.0, $Ls$ = 6.5, $X\!f$ = 7.8, $Y\!f$ = 7.8.}
\label{fig:figura5}
\end{figure}

The circularly polarized antenna is a coax fed squared microstrip patch (Fig. \ref{fig:figura5}). The slot size ($Ws$, $Ls$) and feed position ($X\!f$, $Y\!f$) excite two orthogonal modes that yield circular polarization \cite{Garg2001}. The patch input impedance matching is optimized by varying the feed position along the patch diagonal. 

The frequency bandwidth where polarization remains circular is quite narrow (less than 1\% of the resonant frequency) in this type of antenna but they are very easy to manufacture. The antennas composing the array are rectangular microstrip patches with inset feed (Fig. \ref{fig:figura5}). The patch dimensions ($W\!p$, $Lp$) are obtained from the substrate parameters and the resonance frequency \cite{Balanis2015}. The patch input impedance matching is optimized by varying width and length of the feedline slots ($Si$, $Sg$).
 
\begin{figure}[!t]
\centering
\subfloat[]{\includegraphics[width=8.3cm]{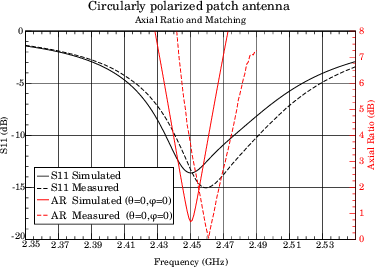}%
\label{fig:figura6a}}
\hspace{0.0cm}
\subfloat[]{\includegraphics[width=8.3cm]{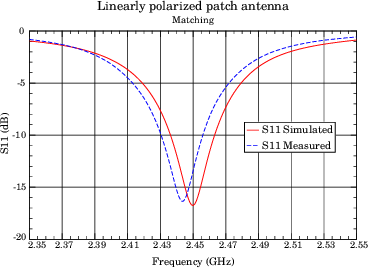}%
\label{fig:figura6b}}
\caption{\textbf{(a)} S11 and Axial Ratio (dB) for circularly polarized antenna and \textbf{(b)} S11 for linearllinºy polarized antenna.}
\label{fig:figura6}
\end{figure}

The substrate used to manufacture the antennas is Rogers RO4360G2 with 1.524 mm height. The relative permittivity of this substrate is quite high (6.15), which results in not too large patch sizes. The antenna size and weight are particularly important in the case of drones because using smaller antennas, wind resistance is reduced and payload efficiency is increased. The Fig. \ref{fig:figura6} shows S11 parameter corresponding to circularly and linearly polarized antennas. Additionally, axial ratio of circularly polarized antenna is included in Fig. \ref{fig:figura6a}.

\subsection{Axial Ratio vs. Drone Location}
The choice of a circularly polarized antenna for the $LP$ avoids amplitude variations in each linearly polarized antenna of the drone as it rotates about its axis. When the Axial Ratio ($AR$) is equal to 1 ($AR$ = 0 dB), the theoretical amplitude variation is zero. However, the $AR$ can vary as a function of the angle (pointing angle) formed by the drone location ($\overline{OL}$ in Fig. \ref{fig:figura1}) and the vertical to the $LP$. This results in a variation of the received amplitude \cite{Balanis2015} and therefore of the maximum drone tracking distance.

\begin{figure}[!t]
\centering
\includegraphics[width=8.5cm]{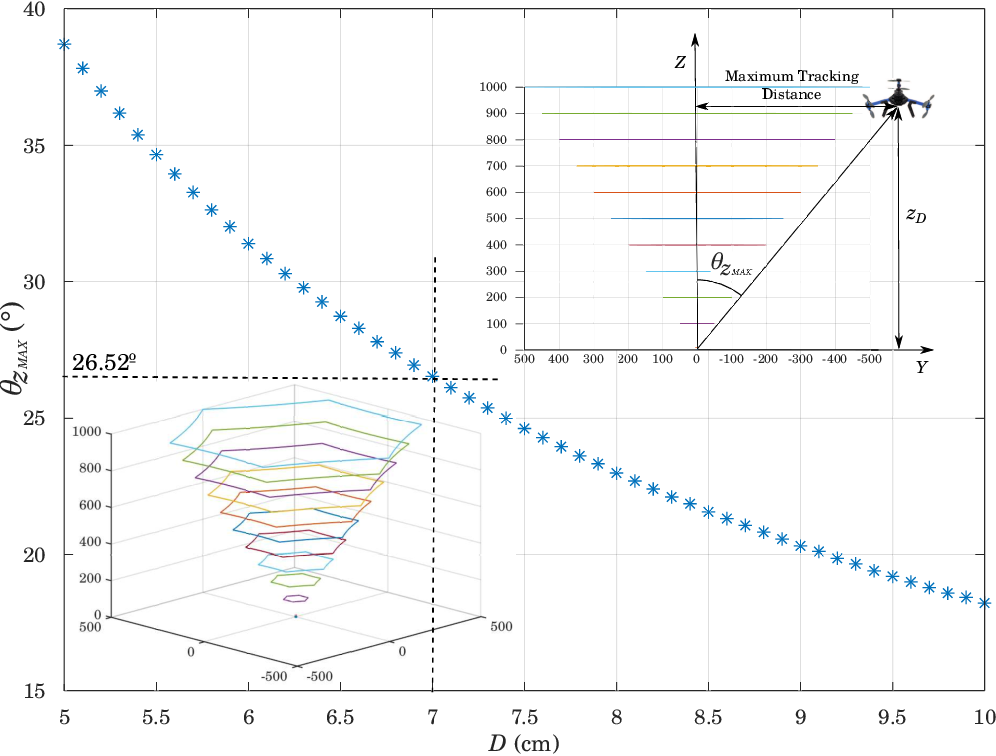}
\caption{Semi-vertical angle of the inverted cone of tracking volume as a function of the distance between receiving antennas ($D$) when f = 2.45 GHz. The tracking volume and its projection in the $ZY$ plane for $D$= 7 cm have been superimposed.}
\label{fig:figura7}
\end{figure}

\begin{figure}[!t]
\centering
\includegraphics[width=8.3cm]{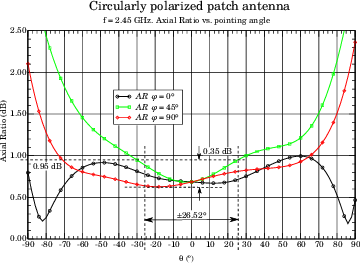}
\caption{Axial ratio ($AR$) of the $LP$ patch antenna for a frequency of 2.45 GHz, as a function of $\theta$ (pointing angle) and for various values of $\varphi$. The range of $AR$ values given by the máximum pointing angle ($\theta_{Z_{MAX}}$ = 26.52\textordmasculine) when $D$ = 7 cm have been marked.}
\label{fig:figura8}
\end{figure}

To see the effect of the pointing angle on the $AR$, several simulations have been performed. As shown in Fig. \ref{fig:figura8}, the range of pointing angles exceeds $\pm$70\textordmasculine\ degrees when the $AR$ is 2 dB. In fact, the maximum pointing angle ($\theta_{Z_{MAX}}$ in Fig. \ref{fig:figura7}) is defined when the drone is farthest from the vertical to the $LP$ (maximum tracking distance), that is, for angles $\phi_L$ multiples of 60\textordmasculine\ (Fig. \ref{fig:figura1b}). The $\theta_{Z_{MAX}}$ angle corresponds to the semi vertical angle of the inverted cone that determines the tracking volume (Fig. \ref{fig:figura7}). The tracking volume has been obtained by superimposing the tracking areas at different drone altitudes (Fig. \ref{fig:figura1}). Fig. \ref{fig:figura7} has plotted the variation of the angle $\theta_{Z_{MAX}}$ as a function of the distance between receiving antennas ($D$) when the frequency is 2.45 GHz. The angle $\theta_{Z_{MAX}}$ is 26.52\textordmasculine\ for the design value ($D$ = 7 cm) which corresponds to a maximum variation of 0.35 dB in the $AR$ (Fig. \ref{fig:figura8}). In view of these results, amplitude variations due to polarization losses will not be significant ($\le$ 0.35 dB in received power) and the phase detector used, AD8302, will not be affected.

\section{Control System}
\begin{figure}[!t]
\centering
\includegraphics[width=8.5cm]{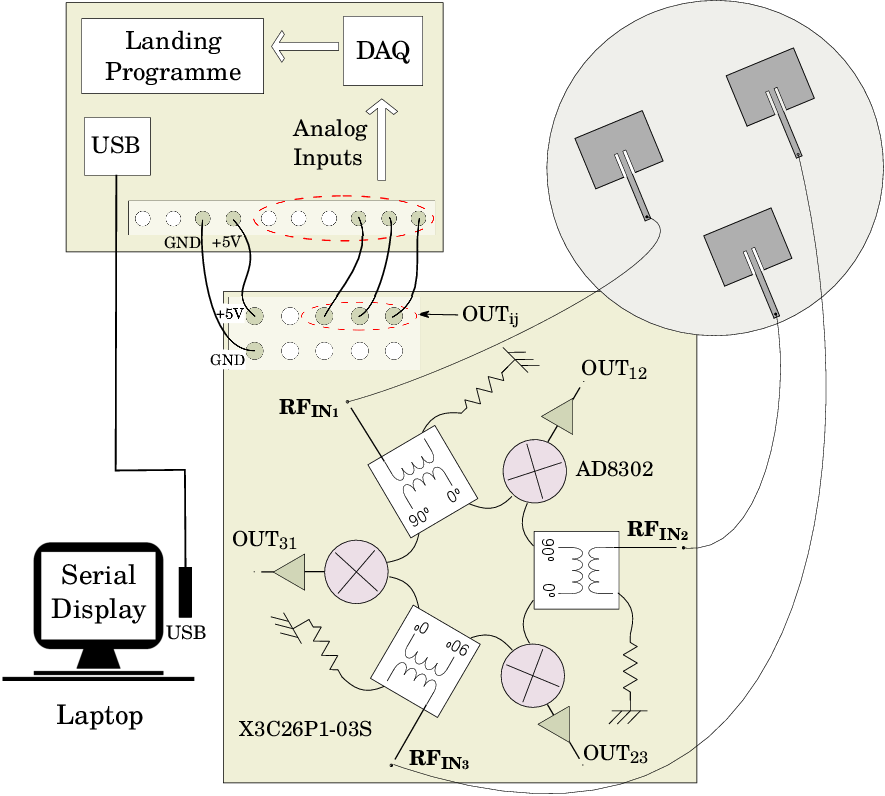}
\caption{Simplified schematic of the drone system assembly.}
\label{fig:figura9}
\end{figure}

The designed drone tri-antenna is connected to a printed circuit board similar to the one developed in \cite{Gautam2014}, whose block diagram is presented in Fig. \ref{fig:figura9}. The schematic also shows the control board (Arduino Uno) that will generate the flight commands associated with phase detector voltages.

The control board is connected via USB to a laptop, which serves to display detected voltages and associated flight instructions generated by the control program. The Arduino Uno has a 10-bit (1024 steps) ADC with an acquisition rate of 10 kHz (100 $\mu$s). The 1024 steps are applied over a maximum voltage of 5 V, providing a resolution of 4.88 mV/step. Since the phase detector provides a 2.6 V maximum voltage variation (0.2 V to 2.8 V), the minimum drone displacement for a level change is 0.746 cm/step (Fig. \ref{fig:figura3}) when the antenna separation is $D$ = 7 cm.

The flow diagram of the control program is shown in Fig. \ref{fig:figura10}. The program starts with the configuration of the input/output ports and the serial port connection speed (USB at 9600 bps) to correctly display the phase detector output voltages and flight maneuvers (turn right/left and move forward/backward). Ten readings are made to calculate averaged phase detector voltage. Measurements are made for each phase detector, which limit the ODR to 333 Hz (3 ms = 0.1 ms\,$\cdot$\,10\,$\cdot$\,3). Next, the zeroing reference voltages (calibration procedure in Section V) are subtracted to match 0\textordmasculine\ phase shift with 0 V voltages ($\pm V_d$ referred to $\pm \Delta \theta$).

\begin{figure}[!t]
\begin{algorithmic}[1]
\small
\STATE \textbf{Initialize}(ports) \COMMENT{Input/Output ports and USB ports to 9600 bps}
\STATE \textbf{Read}(data) \COMMENT{''10 samples for analog detector voltage input``}
\STATE \textbf{Calculate} (''Voltage from Reading Average and zeroing``)
\STATE \textbf{Send} (''Measured and referenced voltage values via USB``)
\STATE \textbf{if} \big(\textbf{abs(${V_d}_{12}$)} $\leq$ \textbf{abs(${V_d}_{23}$)}\big) \& \big(\textbf{abs(${V_d}_{12}$)} $\leq$ \textbf{abs(${V_d}_{31}$)}\big)
\STATE \hspace{0.2cm} \textbf{if} \big(\textbf{${V_d}_{12} \cdot {V_d}_{23} < 0$}\big) \COMMENT{Different Signs}
\STATE \hspace{0.4cm} \textbf{Send} (``Sector 1: Turn LP to Left'')
\STATE \hspace{0.2cm} \textbf{else} 
\STATE \hspace{0.4cm} \textbf{Send} (``Sector 1: Turn LP to Right'')
\STATE \hspace{0.2cm} \textbf{end if}
\STATE \hspace{0.2cm} \textbf{if} \big(\textbf{${V_d}_{23} > 0$}\big) 
\STATE \hspace{0.4cm} \textbf{Send} (``Sector 1b: Forward'')
\STATE \hspace{0.2cm} \textbf{else} 
\STATE \hspace{0.4cm} \textbf{Send} (``Sector 1a: Backward'')
\STATE \hspace{0.2cm} \textbf{end if}
\STATE \textbf{else} 
\STATE \hspace{0.2cm} \textbf{if} \big(\textbf{abs(${V_d}_{23}$)} $<$ \textbf{abs(${V_d}_{31}$)} \big)
\STATE \hspace{0.4cm} \textbf{Send} (``Sector 2: Turn LP to Right 60'')
\STATE \hspace{0.2cm} \textbf{else} 
\STATE \hspace{0.4cm} \textbf{Send} (``Sector 2: Turn LP to Left 60'')
\STATE \hspace{0.2cm} \textbf{end if}
\STATE \textbf{end if} 
\STATE \textbf{Delay(1000)} \COMMENT{Wait 1 seccont for the next iteration}
\end{algorithmic}
\caption{Landing Programme algorithm implemented on Arduino Uno platform.}
\label{fig:figura10}
\end{figure}

The guidance algorithm has been slightly modified to take into account that in the evaluation stage, the $LP$ moves instead of the drone system (Section V). Therefore, the proposed maneuvers are always referred to $LP$ in order to align it with the drone. That is, a movement of the drone towards the $LP$ is contrary to the movement of the $LP$ towards the drone. For example, if the drone is R distance from the $LP$ ($r_L$ = $R$ in Fig. \ref{fig:figura1a}), rotating the drone to the right on its axis is equivalent to moving the $LP$ on $R$ radius circumference in the left direction.

\section{Experimental Evaluation}
\begin{figure}[!t]
\centering
\includegraphics[width=7cm]{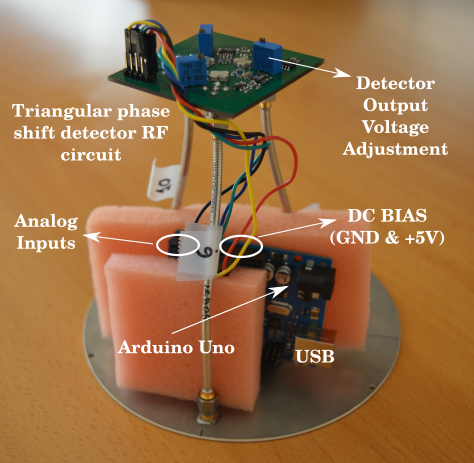}
\caption{Photo of the drone tri-antenna system: tri-antenna, RF board including phase detectors and Arduino Uno control board.}
\label{fig:figura11}
\end{figure}

The drone tri-antenna is connected to the RF board through three balanced wires with a phase difference less than 1\textordmasculine\ at 5 GHz (Fig. \ref{fig:figura11}). Other connections (power supply and phase detector voltages) with the Arduino Uno and laptop have been detailed in Fig. \ref{fig:figura9} schematic. Tests were performed in an indoor enclosure without specific conditioning to reduce possible reflections.

\begin{figure}[!t]
\centering
\subfloat[]{\includegraphics[width=7cm]{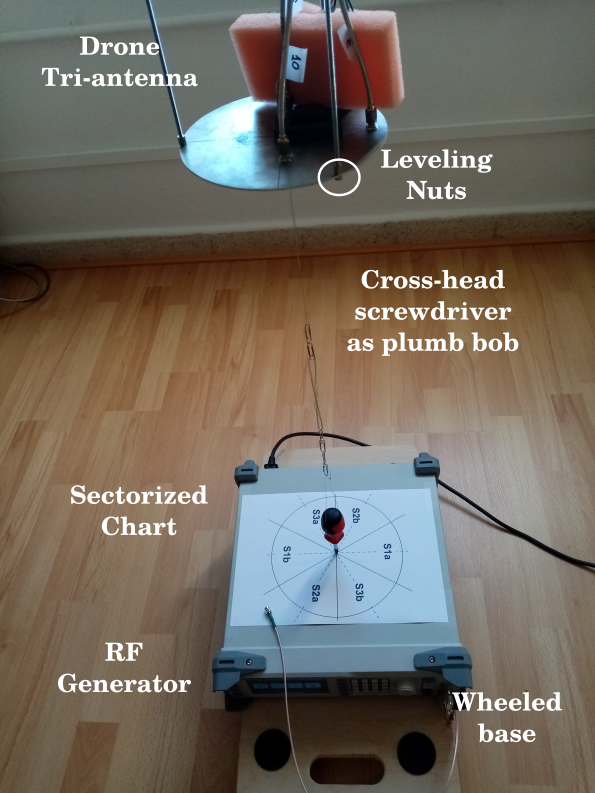}%
\label{fig:figura12a}}
\hspace{0.0cm}
\subfloat[]{\includegraphics[width=7cm]{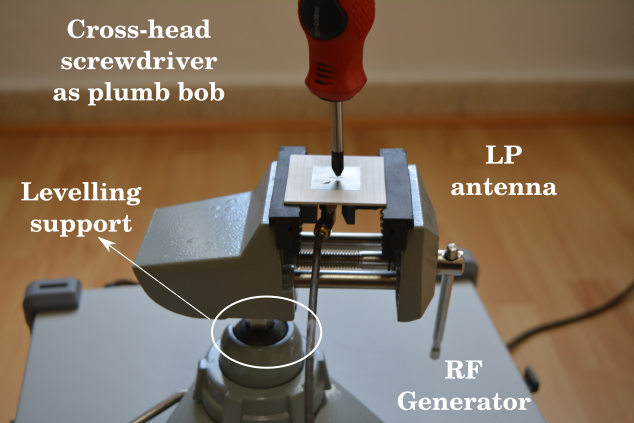}%
\label{fig:figura12b}}
\caption{Photo of the assembly made to check the system performance at short distances. \textbf{(a)} Assembly with $h$ = 46 cm with plumb line (small screwdriver) for Sectorized Chart adjustment. \textbf{(b)} Leveling support used with $LP$ antenna.}
\label{fig:figura12}
\end{figure}

The drone tri-antenna was attached to a wooden support and leveled with the adjustment screws located on the edge of the tri-antenna substrate (Fig. \ref{fig:figura12a}). An RF generator (2.46 GHz and 0 dBm) placed on a wheeled base and connected to the circularly polarized patch antenna was used for $LP$ (Fig. \ref{fig:figura12b}). A leveling support was employed to compensate for potential floor irregularities. 

In the first test, the drone tri-antenna was 46 cm from the $LP$ ($z_D$ = 46 cm), which corresponds to a minimum/maximum tracking distance of 19.3/22.8 cm, considering a maximum phase shift range of $\pm$80\textordmasculine\ (Fig. \ref{fig:figura1b}). A sectorized chart was stuck to the RF generator for reference purposes when moving the $LP$ leveling support. To obtain the zeroing reference voltages, a plumb line was used to place the drone tri-antenna incenter above the perpendicular to the sectorized chart center and $LP$. These values (1.378 V, 1.324 V, 1.336 V) were subtracted from phase detector output voltages in the control program to obtain 0 V when the drone tri-antenna was just above the $LP$ antenna ($\Delta \theta_{12}$ = $\Delta \theta_{23}$ = $\Delta \theta_{31}$ = 0).

Furthermore, in order to facilitate manual positioning and testing, the control program displays the word ``LOCK'' when the voltage is within $\pm$0.1 V range. This simple experimental procedure compensates for possible phase errors arising from connections to drone tri-antenna and mismatching. Once the center was calibrated, the sectorized chart had to be correctly aligned, i.e. the S1a sector bisector had to coincide with the $OP_3$ line (Fig \ref{fig:figura1a}). For this purpose, the $LP$ antenna was moved along the $S1a$ sector bisector and the wheeled base was rotated until 0 V was obtained at ${V_d}_{12}$, i.e., same distance between $LP$ and the receiving points $1$ and $2$. Due to the available resources, center adjustment and rotation to align the sectorized chart had to be performed iteratively until the objective was achieved.

\begin{figure}[!t]
\centering
\includegraphics[width=6.5cm]{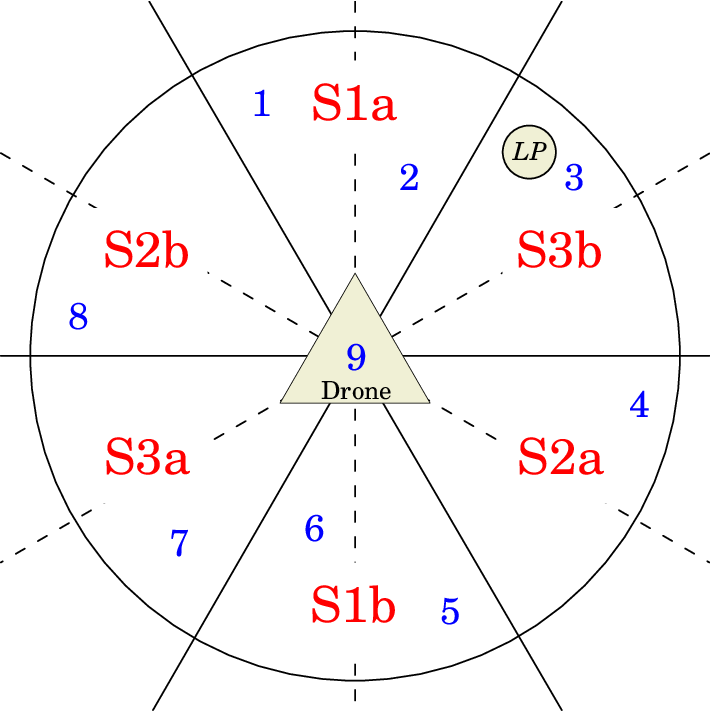}
\resizebox{0.5\textwidth}{!}
{
\begin{tabular}{|c|c|c|c|c|c c c|}
\hline
\textbf{N\textordmasculine} & \textbf{Sector} &  
\parbox{1cm}{\centering $\mathbf{{V_d}_{12}}$ \\ \textbf{(V)}} & 
\parbox{1cm}{\centering $\mathbf{{V_d}_{23}}$ \\ \textbf{(V)}} & 
\parbox{1cm}{\centering $\mathbf{{V_d}_{31}}$ \\ \textbf{(V)}} & & \textbf{Message} & \\ \hline
\textcolor{blue}{1} & \textcolor{red}{S1a} & -0.18 & -0.45 & 0.62 & Turn LP & to Right & Backward\\ \hline
\textcolor{blue}{2} & \textcolor{red}{S1a} & 0.45 & -0.92 & 0.48 & Turn LP & to Left & Backward\\ \hline
\textcolor{blue}{3} & \textcolor{red}{S3b} & 0.9 & -0.76 & -0.15 & Turn LP &  to Left & 60 Deg\\ \hline
\textcolor{blue}{4} & \textcolor{red}{S2a} & 0.6 & 0.23 & -0.83 & Turn LP & to Right & 60 Deg\\ \hline
\textcolor{blue}{5} & \textcolor{red}{S1b} & 0.37 & 0.62 & -1.0 & Turn LP & to Right & Forward\\ \hline
\textcolor{blue}{6} & \textcolor{red}{S1b} & -0.17 & 0.66 & -0.51 & Turn LP & to Left & Forward\\ \hline
\textcolor{blue}{7} & \textcolor{red}{S3a} & -0.5 & 0.43 & 0.07 & Turn LP & to Left & 60 Deg\\ \hline
\textcolor{blue}{8} & \textcolor{red}{S2b} & -0.56 & -0.25 & 0.8 & Turn LP & to Right & 60 Deg\\ \hline
\textcolor{blue}{9} & \textcolor{red}{Center} & -0.02 & -0.04 & 0.05 & Lock & Lock & Lock\\
\hline
\end{tabular}
}
\caption{Phase detector voltage measurements corresponding to each sector ($z_D$ = 46 cm). The LOCK state is considered for voltages within $\pm$0.1 V.}
\label{fig:figura13}
\end{figure}

Fig. \ref{fig:figura13} shows the phase detector voltages corresponding to different $LP$ locations in each sector. The voltage values associated with each sector generate expected flight correction messages, referred to $LP$ movements in this case.

Then, the wheeled base was moved with the $LP$ antenna centered on sectorized chart until a maximum voltage ${V_d}_{12}$ was obtained, i.e., the voltage that delimits the non ambiguous tracking area. A voltage of 1.06 V was obtained when the displacement was about 21 cm, close to the theoretical value of 23 cm (Fig. \ref{fig:figura1b}). The system was tested for $z_D$ = 31 cm, with tracking area 13/15 cm (14 cm measured) and $z_D$ = 16 cm with tracking area 7/8 cm (7 cm measured). In addition, a test was performed at 1.8 cm height (near field condition \cite{Balanis2015}) where it was only possible to move $LP$ according to flight instructions until LOCK was achieved because the size of tracking area was too small ($<$0.7 cm).

\begin{figure}[!t]
\centering
\subfloat[]{\includegraphics[width=2.65cm]{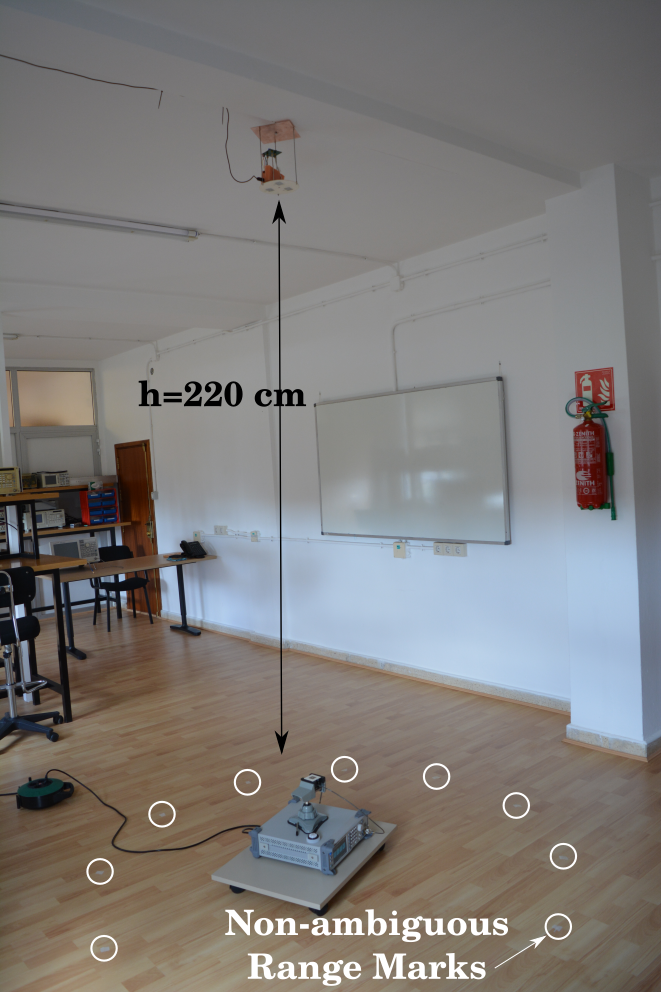}
\label{fig:figura14a}}
\hspace{0.0cm}
\subfloat[]{\includegraphics[width=5.9cm]{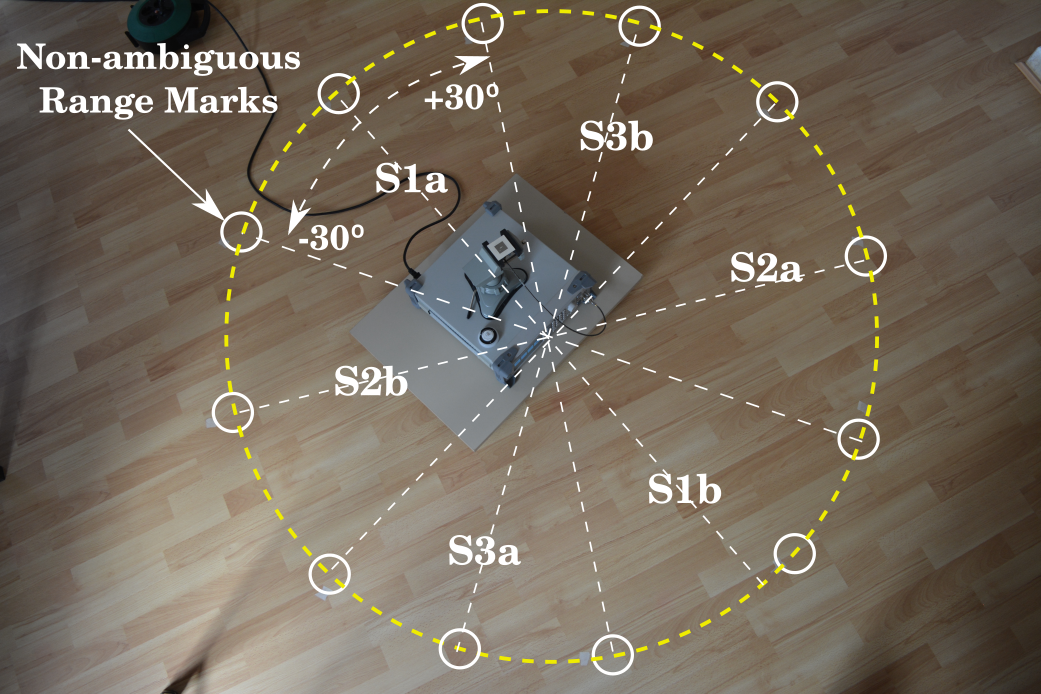}%
\label{fig:figura14b}}
\caption{Photo of the assembly made to check the operation of the system at $z_D$ = 220 cm height. Points delimiting a radius of about 92 cm and tracking area sectors have been marked.}
\label{fig:figura14}
\end{figure}

In the next test, the drone tri-antenna was attached to the roof of the room, at 220 cm height (far field condition \cite{Balanis2015}) from the $LP$ antenna (Fig. \ref{fig:figura14a}). At this height, there was a minimum/maximum tracking distance of 92/110 cm (Fig. \ref{fig:figura1b}). In Fig. \ref{fig:figura14b}, a set of marks was superimposed to delimit the tracking area of about 92 cm radius. Further, corresponding sectors were superimposed, according to drone tri-antenna orientation and measurements. The wheeled base was moved on ground, checking that the instructions guide $LP$ to the drone tri-antenna center.

Finally, an amplitude dynamic range of 28 dB was obtained (green in Table \ref{tab:table_i}) by varying the RF power and measuring detected voltages until a maximum permissible error of $\pm$0.1 V (LOCK) was exceeded when the drone tri-antenna was just above $LP$. The upper range was limited by the maximum power provided by the RF generator (20 dBm).

\begin{table}[!t]
\caption{Detected voltages when the drone tri-antenna is just above the $LP$ antenna and the transmitted power is varied.}
\label{tab:table_i}
\centering
\begin{tabular}{|c|c|c|c|}
\hline
\textbf{RF Power} & $\mathbf{\Delta {V_d}_{12}}$ & $\mathbf{\Delta {V_d}_{23}}$ & $\mathbf{\Delta {V_d}_{31}}$ \\ 
\textbf{(dBm)} & \textbf{(V)} & \textbf{(V)} & \textbf{(V)} \\ \hline
\rowcolor{color1} -15 & -0.16 & 0.22 & 0.38 \\ \hline
\rowcolor{color1} -10 & -0.12 & 0.11 & 0.09 \\ \hline
\rowcolor{color1} -9 & -0.11 & 0.11 & 0.08 \\ \hline
\rowcolor{color2} -8 & -0.02 & 0.06 & 0.03 \\ \hline
\rowcolor{color2} -7 & -0.07 & 0.01 & 0.03 \\ \hline
\rowcolor{color2} -6 & -0.05 & -0.01 & 0.01 \\ \hline
\rowcolor{color2} -5 & 0.02 & 0 & -0.03 \\ \hline
0 & 0 & 0 & 0 \\ \hline
\rowcolor{color2} +5 & 0.03 & 0.01 & -0.01 \\ \hline
\rowcolor{color2} +10 & -0.01 & 0.01 & -0.01 \\ \hline
\rowcolor{color2} +15 & -0.04 & 0.08 & -0.02 \\ \hline
\rowcolor{color2} +20 & 0.04 & 0.01 & -0.04 \\
\hline
\end{tabular}
\end{table}

\section{Conclusions}
A radio frequency instrument used for precision vertical landings has been designed. The authors have started from a procedure proposed in \cite{Arana2021} and a circuit provided with three RF inputs, which was evaluated from signals from a set of phase-synchronized generators. The flight correction instructions (forward/backward and right/left) for steering towards the landing point are obtained by detecting phase shifts between RF input signals.

The necessary elements to operate in signal propagation conditions (drone tri-antenna, landing point antenna, and control circuit) are added in this work, determining the design parameters and experimentally evaluating their behavior. As a result, the constructed system shows proper functioning in far and near field conditions and provides a higher ODR than existing systems. The most relevant contributions are detailed below. The equations governing the tracking area and the characteristics of the phase detector and ADC provide the antenna system design: distance between the drone tri-antenna; tracking distance; sensitivity defined as phase shift voltage versus drone flying distance in V/cm; and ADC uncertainty in cm/step, related to drone flying distance without a change in ADC level. Moreover, to avoid amplitude variations in the input signals when the drone turns on its axis, the landing point antenna has been designed to be circularly polarized. The drone tri-antenna system was designed with linear polarization and the same orientation to ensure that phase shifts between antennas have the same angular reference. The manufactured prototype uses patch antennas, a simple Arduino Uno platform controller with a 10-bit ADC (0.1 ms per sample), and a commercial phase detector (AD8302) that provides an output voltage of 2.6 V (0.2 V to 2.8 V) for an input phase shift of $\pm$80\textordmasculine\ at 2.46 GHz. The drone tri-antenna was designed with a 7 cm separation between antennas, which provides a maximum tracking area of 499 cm at 10 m, a sensitivity of 2.605 mV/cm and a maneuvering uncertainty due to the ADC resolution of 0.746 cm/step. Additionally, the amplitude variation of the input signal due to the $AR$ variation of the circularly polarized $LP$ antenna has been quantified. The maximum variation is 0.35 dB, which corresponds to an angle of $\pm$26.52\textordmasculine\ when the drone is at the border of the tracking volume (inverted cone).

The prototype showed a robust performance according to the design, despite the multiple reflections that exist in the room used.
The most significant results obtained from the measurements are the following: ODR ($>$300 Hz) greater than available systems for landing maneuvers,  dynamic range close to 30 dB (RF power transmitted between -8 and 20 dBm for a distance between antennas of 220 cm), coherence in the flight correction orders in the tests performed between 220 cm (far field condition \cite{Balanis2015}) and 1.8cm (near field condition \cite{Balanis2015}) with 0 dBm RF power transmitted. Finally, a calibration procedure using voltage zeroing detector has been proposed, which avoids the use of external RF generators to calibrate the phase shift system and absorbs the problems of mismatching and characteristics of the used antennas (matching and connection cables).

\section*{Acknowledgments}
Thanks to Juan Domingo Santana Urb\'in for his enormous contribution during this research. This work was supported by the Spanish Gobernment under Grant (TEC2017-88242-C3-3-R) and (PID2020-116569RB-C32) Projects.


\bibliographystyle{IEEEtran}
\bibliography{IEEEabrv,mybibliography.bib}

%
\vspace{-1cm}
\begin{IEEEbiography}[{\includegraphics[width=1in,height=1.25in,clip,keepaspectratio]{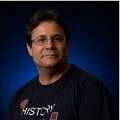}}]{V\'ictor Ara\~na-Pulido} (Member, IEEE) was born in Las Palmas, Spain, in 1965. He received the M.Sc. degree from the Universidad Polit\'ecnica de Madrid (UPM), Madrid, Spain, in 1990, and the Ph.D. degree from the Universidad de Las Palmas de Gran Canaria (ULPGC), Las Palmas, in 2004. He is currently an Assistant Professor with the Signal and Communication Department and a member of the Institute for Technological Development and Innovation in Communications (IDeTIC), ULPGC. His current research interests include the nonlinear design of microwave circuits, control subsystem units, and communications systems applied to data acquisition complex networks.
\end{IEEEbiography}

\vspace{-1cm}
\begin{IEEEbiography}[{\includegraphics[width=1in,height=1.25in,clip,keepaspectratio]{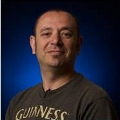}}]{Eugenio Jim\'enez-Ygu\'acel} (Member, IEEE) received the M.Sc. and Ph.D. degrees from the Universidad Polit\'ecnica de Madrid (UPM), Madrid, Spain, in 1991 and 2002, respectively. He joined the University of Las Palmas de Gran Canaria, Las Palmas, Spain, in 1993, where he is currently an Associate Professor and a member of Institute for Technological Development and Innovation in Communications (IDeTIC-ULPGC). His initial activity focuses on the design, manufacture, measurement of antennas and microwave circuits. Along these lines, he has participated in national public projects uninterruptedly since 1988. In 2006, he began a new line related to digital communications in HF (1–30 MHz) and signal processing through programmable logic.
\end{IEEEbiography}

\vspace{-1cm}
\begin{IEEEbiography}[{\includegraphics[width=1in,height=1.25in,clip,keepaspectratio]{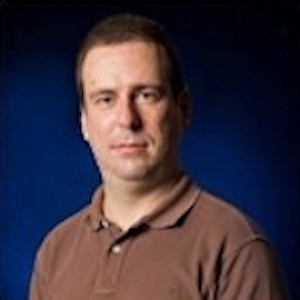}}]{Francisco Cabrera-Almeida}  (Member, IEEE) was born in Las Palmas, Spain, in 1970. He received the M.Sc. and Ph.D. degrees from the Universidad de Las Palmas de Gran Canaria (ULPGC), Las Palmas, in 1997 and 2012, respectively. He is currently an Assistant Professor with the Signal and Communications Department and a member of the Institute for Technological Development and Innovation in Communications (IDeTIC), ULPGC, since it was founded in 2010. He has taken part in a number of Spanish and European projects in collaboration with industries and other universities. His current research interests include numerical electromagnetic modeling techniques and radiowave propagation and communications systems applied to data acquisition complex networks.
\end{IEEEbiography}

\vspace{-1cm}
\begin{IEEEbiography}[{\includegraphics[width=1in,height=1.25in,clip,keepaspectratio]{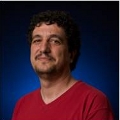}}]{Pedro Quintana-Morales} was born in Santa Cruz de Tenerife, Spain, in 1964. He received the M.Sc. degree from the Universidad Polit\'ecnica de Madrid (UPM), Madrid, Spain, in 1989, and the Ph.D. degree from the Universidad de Las Palmas de Gran Canaria (ULPGC), Las Palmas, Spain, in 2016. He is currently an Assistant Professor with the Signal and Communication Department and a member of Institute for Technological Development and Innovation in Communications (IDeTIC-ULPGC). He has taken part in a number of Spanish and European projects in collaboration with industries and other universities. His current research interests include signal processing, data analysis applied to speech, images, biosignals, and radio communications and sensor networks.
\end{IEEEbiography}

\vfill

\end{document}